\documentclass[twocolumn]{aastex63}
\hypersetup{linkcolor=red,citecolor=green,filecolor=cyan,urlcolor=magenta}
\usepackage{booktabs}
\usepackage{bookmark}
\usepackage{xspace}

\newcommand{\tess}{{\em TESS}\xspace}

\newcommand{\keV}{{\rm ~keV}\xspace}

\newcommand{\s}{{\rm ~s}\xspace}
\newcommand{\hr}{{\rm ~hr}\xspace}
\newcommand{\m}{{\rm ~min}\xspace}
\newcommand{\ms}{{\rm ~ms}\xspace}
\newcommand{\muhz}{{\rm ~$\mu$Hz}\xspace}
\newcommand{\mupo}{{\rm ~$\frac{e^-}{\mu Hzs^2}$}\xspace}
\newcommand{\eps}{{\rm ~${e^-}~s^{-1}$}\xspace}
\newcommand{\fluxcgs}{{\em ergs~s$^{-1}$~cm$^{-2}$}\xspace}

\newcommand{\psroz}{{PSR~J1023+0038}\xspace}
\newcommand{\psrof}{{PSR~J1417-4402}\xspace}
\newcommand{\gltwo}{{3FGL~J0212.1+5320}\xspace}
\newcommand{\glfiveold}{{1FGL~J0523.5-2529}\xspace}
\newcommand{\glfive}{{3FGL~J0523.3-2528}\xspace}
\newcommand{\glseven}{{3FGL~J0744.1-2523}\xspace}
\usepackage{color}
\usepackage{amsmath}

\shorttitle{Periodicity Search with \tess}
\shortauthors{Pal et al.}

\begin{document}

\title{Periodicity search for Pulsar Binaries with \tess}

\author[0000-0001-8922-8391]{Partha Sarathi Pal}
\affiliation{School of Physics and Astronomy, Sun Yat Sen University, Guangzhou 510275, China}
\author[0000-0002-1262-7375]{P.~H.~T.~Tam}
\affiliation{School of Physics and Astronomy, Sun Yat Sen University, Guangzhou 510275, China}
\author[0000-0003-3773-5302]{Weitang Liang}
\affiliation{School of Physics and Astronomy, Sun Yat Sen University, Guangzhou 510275, China}
\author[0000-0001-6655-854X]{Chengye Cao}
\affiliation{School of Physics and Astronomy, Sun Yat Sen University, Guangzhou 510275, China}
\author[0000-0002-0439-7047]{K.~L.~Li}
\affiliation{Institute of Astronomy, National Tsing Hua University, Hsinchu 30013, Taiwan}
\author[0000-0003-1753-1660]{C.~Y.~Hui}
\affiliation{Department of Astronomy \& Space Science, Chungnam National University, Daejeon 34134, Korea}
\author[0000-0002-5105-344X]{A.~K.~H.~Kong}
\affiliation{Institute of Astronomy, National Tsing Hua University, Hsinchu 30013, Taiwan}
\correspondingauthor{P.~H.~T.~Tam}
\email{tanbxuan@mail.sysu.edu.cn}

\begin{abstract}
Pulsar binaries, in particular redback systems, provide good 
sources to study the pulsar wind flow and its interaction with 
the companion stars. {\it Fermi}-LAT have proposed probable 
pulsar binary candidates in its catalogs. To identify pulsar binary 
sources from the catalog, orbital modulation search of binary 
candidates is an effective way. \tess observes 
in survey mode for a large part of the sky and thus provide an 
excellent data set to periodicity search of pulsar binary candidates 
by observing the flux variation, thought to mainly come from the 
stellar companion. Using \tess data we look for flux modulation 
of five pulsar binaries (or candidates) with reported orbital periods, 
including PSR J1023+0038, 3FGL J0523.3-2528, 3FGL J0212.1+5320, 
3FGL J0744.1-2523 and PSR J1417-4402, demonstrating that \tess 
photometric data are very useful in identifying periodicities 
of redback-like systems. This method can be effective in searches for 
new pulsar binaries or similar binary systems in the future.  
\end{abstract}
\keywords{methods: data analysis --- pulsars: general --- X-rays: binaries --- (stars:) pulsars: individual (PSR J1023+0038, 3FGL J0523.3-2528, 3FGL J0212.1+5320, 3FGL J0744.1-2523, PSR J1417-4402)}

\section{Introduction} \label{sec:intro}

Redback systems are close-orbit pulsar binaries that show intense 
interactions between the pulsars and the companion stars. 
The orbital period of redbacks typically spans $P_\mathrm{b}\leq\,20$~hrs 
\citep{Roberts2013,Hui19}, with an exception 
PSR~J1306-40 $P_\mathrm{b}=26.3$~hrs \citep{linares18,swihart19}.
The companion stars are late-type non-degenerate ones with 
masses of $M_\mathrm{c}\sim0.2 - 0.5\,M_{\sun}$, 
which are significantly higher than those of another similar 
class of systems, black widows, for which 
$M_\mathrm{c}<0.1\,M_{\sun}$ \citep{Roberts2013,Hui19}. 
Some redbacks have been observed to transit between 
the rotational-powered state and a state with an accretion disk, 
such as those observed in \psroz and XSS~J12270-4859 
\citep[see, e.g.][and references therein]{arch09,papitto13,patruno14,stappers14,bassa14,takata14,roy15}. Pulsar binaries of longer orbital periods than one day, such as \psrof, have also been discovered \citep[e.g.][]{swihart18}.

Optical periodicity related to the pulsar irradiation and 
ellipsoidal variation have been observed in many of the known pulsar binaries. 
At the same time, non-periodic phenomena such as optical flares 
(e.g., from the accretion disk) 
and flux change (between different states/modes) at various time scales 
can also be used to probe the astrophysical conditions of the systems.

{\it Transiting Exoplanet Survey Satellite (TESS)} is a survey satellite with a bandpass of $600-1000$ nm
whose principal mission is to observe flux variation of stars 
to determine the presence of exoplanets around those stars \citep{tess15}. 
With its supreme timing ability at time 
scales of minutes to hours and its nearly all-sky coverage, it is also 
an ideal instrument to characterize periodicities 
and flux variations of stars \citep{dorn19,balona20}
and binaries, such as redbacks either in their accretion or 
rotation-powered states.

In this work, we searched through the literature for known 
redback-like pulsar binaries (or candidates) with reported 
orbital period, including those covered by \tess target products. 
The chosen sources are given in Table.~\ref{src_list}. The aim is to 
demonstrate the photometric capabilities of \tess to characterize 
redbacks and similar systems.

\begin{deluxetable*}{ccccccccc}
\tablenum{1}
\tablecaption{List of sources analysed: (1)~Source names, (2)~Right Accesion, (3)~Declination, (4)~Observation start time, (5)~Observation stop time, (6)~Sector, (7)~Camera No., (8)~CCD No., (9)~\tess magnitude.
\label{src_list}}
\tabletypesize{\scriptsize}
\tablewidth{0pt}
\tablehead{
\colhead{Source Name} &  \colhead{RA} & \colhead{Dec} & \colhead{Start time} & \colhead{Stop time} & \colhead{Sector} & \colhead{Camera} & \colhead{CCD} & \colhead{\tess} \\
\colhead{} &  \colhead{($\arcdeg$)} & \colhead{($\arcdeg$)} & \colhead{(UTC)} & \colhead{(UTC)} & \colhead{(\#)} & \colhead{(\#)} & \colhead{(\#)} & \colhead{(Mag.)} \\
}
\colnumbers
\startdata
PSR~J1023+0038 & 155.948668 & 0.644794 & 2019-02-02 20:09:36 & 2019-02-27 11:59:35 & 8 & 1 & 3 & 16.28 \\
3FGL~J0523.3-2528 & 80.820517 & -25.460263 & 2018-11-15 11:25:40 & 2018-12-11 18:53:39 & 5 & 2 & 2 & 15.78 \\  
{} & {} & {} & 2018-12-15 18:27:39 & 2019-01-06 13:03:39 & 6 & 2 & 1 & 15.78 \\ 
3FGL~J0212.1+5320 & 33.043655 & 53.360771 & 2019-11-03 03:35:26 & 2019-11-27 12:43:25 & 18 & 2 & 3 & 13.85 \\ 
3FGL~J0744.1-2523 &  116.044700 & -25.399400 & 2019-01-08 02:59:37 & 2019-02-01 13:59:36 & 7 & 2 & 2 & \nodata \\
PSR~J1417-4402 &  214.377517 & -44.049269 & 2019-04-23 06:29:33 & 2019-05-20 08:59:32 & 11 & 1 & 2 & \nodata \\
\enddata
\end{deluxetable*}

\section{The redback systems (candidates)} \label{sec:sample}
\subsection{\psroz}
The prototypical redback pulsar, \psroz, has shown transitions 
between a LMXB state and a rotational-powered 
state~\citep[see, e.g.,][]{arch09,tam10,patruno14,stappers14}.
A single-humped modulation in optical for 4.75\hr was first reported 
in \citet{woudt04}, during the time now believed to be a pulsar state. 
Time-resolved optical spectroscopy and 
photometry of \psroz revealed that it is an X-ray binary and consists 
of a late type G5 companion star with a period of 4.75\hr \citep{ta05}.
Subsequently, detection of a pulsar spin period of 1.69\ms 
in 2007 confirmed the primary as a radio millisecond pulsar \citep{arch09}.
Using further observations, it has been found that the donor 
star has a mass $M2 \sim 0.24M_{\odot}$, $M_{NS} = 1.71 \pm 0.16 M_{\odot}$, 
and the binary at a distance $d = 1.37 \pm 0.04~kpc$ 
\citep{deller12, mcconnell15}.
In 2013 the transition from MSP to LMXB state of \psroz was 
reported with disappearance of radio pulsation and increase 
in optical, X-ray and $\gamma$-ray fluxes 
\citep{patruno14, stappers14, takata14}.

Since then, in the current accretion state, 
\psroz shows rapid flickering and double-peaked emission lines in 
a blue optical spectrum, believed to be associated with an 
accretion disc \citep[e.g.][]{kennedy18,shahbaz19}, 
as first seen in the previous accretion state \citep{szkody03}.

On {\it 2019-02-02 20:09:35 UTC} \tess has observed \psroz for 
27 days under \tess GI Proposal id: \#G011187 (PI: Mark Kennedy).

\subsection{\glfive}
\glfiveold was discovered as a {\it Fermi-LAT} unidentified 
$\gamma$-ray source \citep{abdo10}, without detected radio 
emission yet \citep{guil12, petrov13}. Later it was recataloged 
as \glfive \citep{acero15}. The optical photometry 
and {\it SOAR} spectroscopic observations of a X-ray source 
detected within the localization error of \glfive revealed a 
periodic flux modulation of 16.5\hr period~\citep{strader14}. 
The radial velocity 
variations indicate a probable binary pulsar with an unusually 
massive ($0.8~M_{\odot}$) secondary companion and a measurable 
eccentricity (e = 0.04)\citep{strader14}.

On {\it 2018-11-15 11:25:39 UTC, 2018-12-15 18:27:39 UTC}; \tess 
has observed \glfive for 54 days under \tess GI Proposal id: 
\#G011187 (PI: Mark Kennedy).

\subsection{\gltwo}
\gltwo was first discovered as an unidentified $\gamma$-ray source, 
$1FGL~J0212.3+5319$ \citep{abdo10}. Detailed photometry 
and optical spectroscopy 
classified \gltwo as a redback MSP candidate with a 
period of 0.87~days \citep{li16, linares17}. From spectroscopic 
modeling of \gltwo optical data, it is reported that \gltwo binary 
system may consists of a neutron star and secondary star mass of 
$M1 = 1.85^{+0.32}_{-0.26} M_{\odot}$ and 
$M2 = 0.50^{+0.22}_{-0.19} M_{\odot}$ respectively \citep{shahbaz17}. 

On {\it 2019-11-03 03:35:25 UTC}, \tess has observed \gltwo for 
27 days exposure time under \tess GI Proposal id: \#G022055 
(PI: Francesco Coti Zelati).

\subsection{\glseven}
\glseven was detected as an unidentified $\gamma$-ray source. 
No associated X-ray source, within 0.3-10 \keV 3-$\sigma$ upper 
limit, is detected down to a limit of $4.5 \times 10^{-14}$ \fluxcgs .
The field of \glseven is not covered by the Catalina Sky Survey. 
A variable {\it GROND} source within the error ellipse of \glseven 
has been found. This source features a clear 
flux modulation with an optical and near-IR period equal to 
0.115~day \citep{salvetti17}. We note that such a candidate 
optical counterpart is now outside the 4FGL error circle, and 
no bright radio or optical counterpart can be found within the 
4FGL 95\% error circle.

On {\it 2019-01-08 02:59:36 UTC}, \tess has observed \glseven 
sky location during Sector: \#7 of its survey mode for 
27 days exposure time. 
 
\subsection{\psrof}
\psrof was detected as a $\gamma$-ray source by {\it Fermi-LAT} 
and cataloged as 3FGL~J1417.5-4402 \citep{abdo10}.
Photometric and spectroscopic analysis of the optical counterpart 
reported a period of 5.37 day, with no significant eccentricity, 
at a distance of 4.4 kpc. The mass ratio of the system is predicted 
around $\frac{M2}{M_{NS}}=0.18$. The estimated mass of the components 
are $M_{NS} = 1.97 \pm 0.15 M_{\odot}$ and $M2 = 0.35 \pm 0.04 M_{\odot}$ 
\citep{strader15}. A 2.66\ms radio pulsar \psrof has been found and 
the distance from radio data is estimated to be 1.6 kpc \citep{camilo16}. 
Subsequently, \psrof is classified to be a redback-like system 
\citep{swihart18, devito19}. 

On {\it 2019-04-23 06:29:33 UTC}, \tess 
has observed \psrof sky location during Sector: \#11 of its 
survey mode for 27 days exposure time.

\section{Data Analysis}

Periodicity searches were performed using archival 
\tess data. The obtained periodicity is then compared to the  
reported orbital periods.

\begin{figure*}[ht]
\centering
\gridline{
\fig{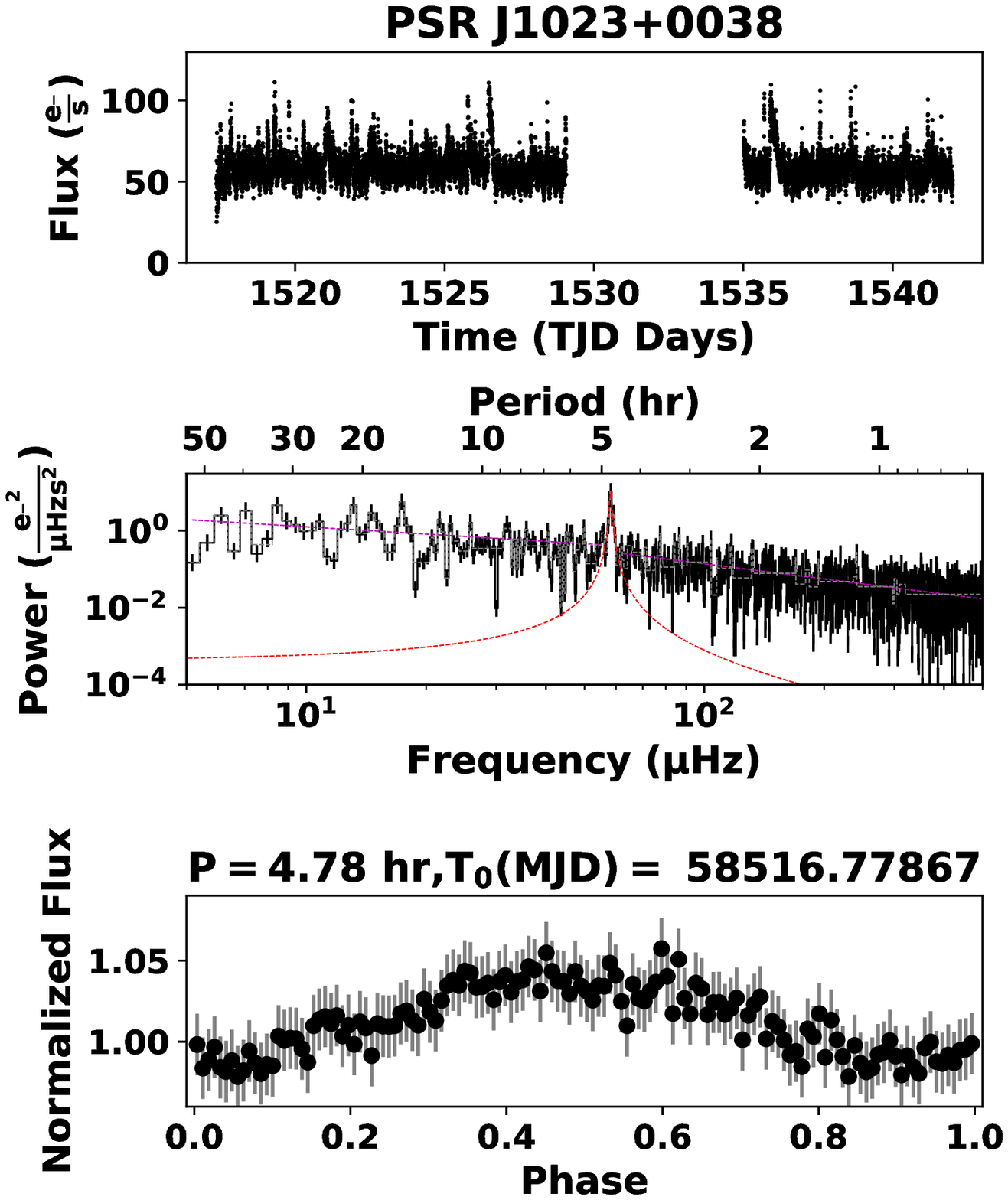}{0.48\textwidth}{(a):~Analysis result of \psroz}
\fig{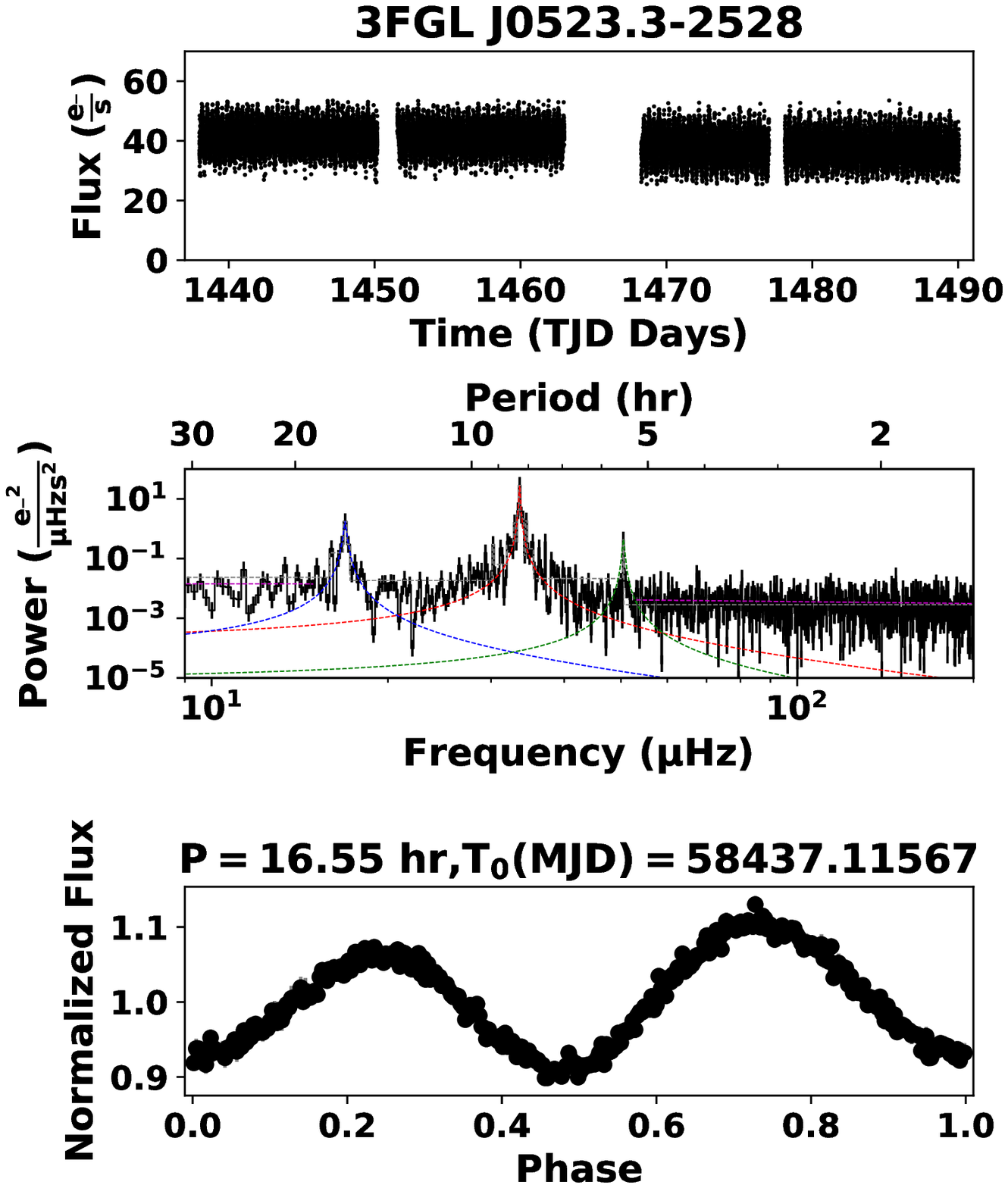}{0.48\textwidth}{(b):~Analysis result of \glfive}}
\gridline{
\fig{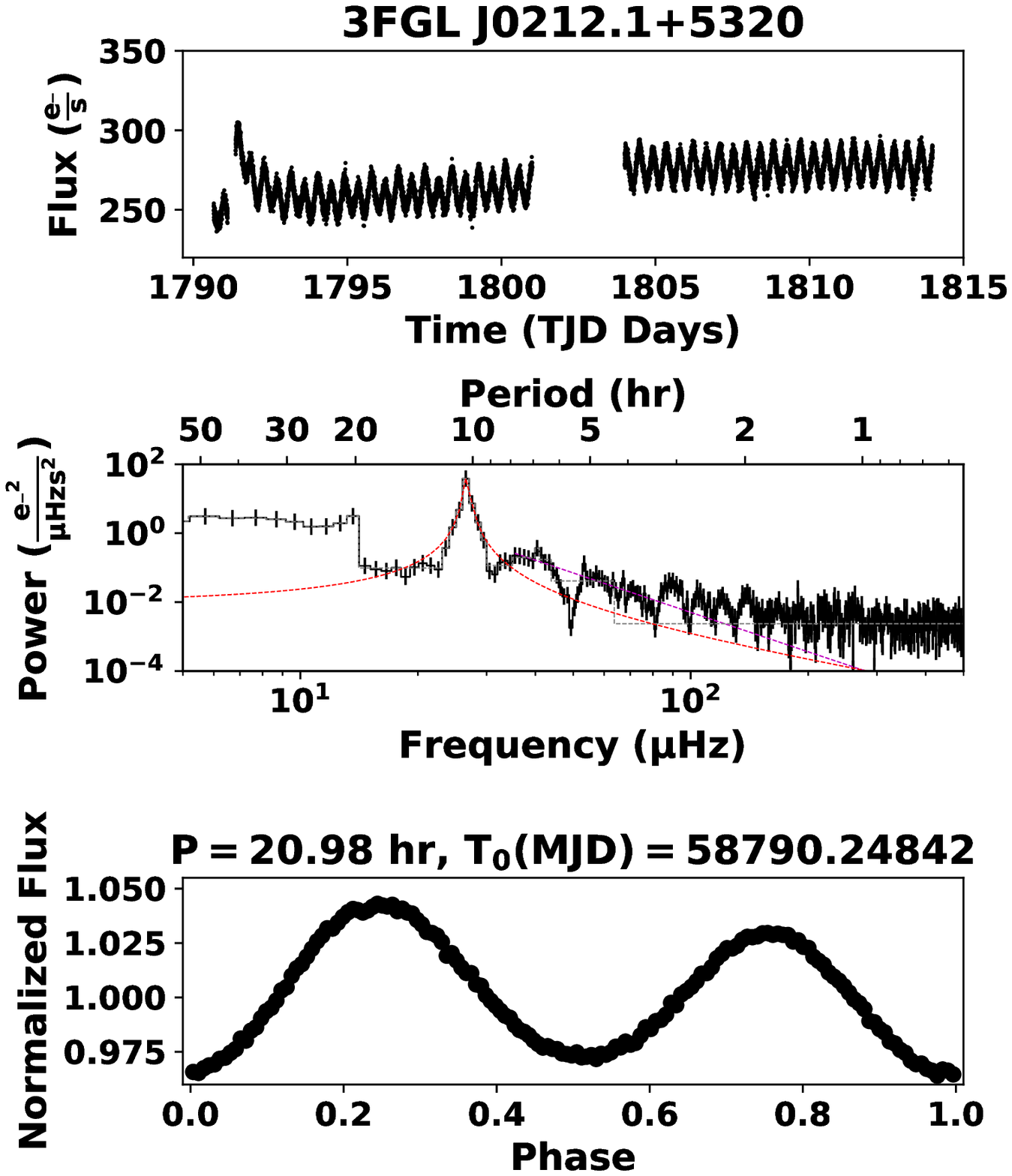}{0.48\textwidth}{(c):~Analysis result of \gltwo}
\fig{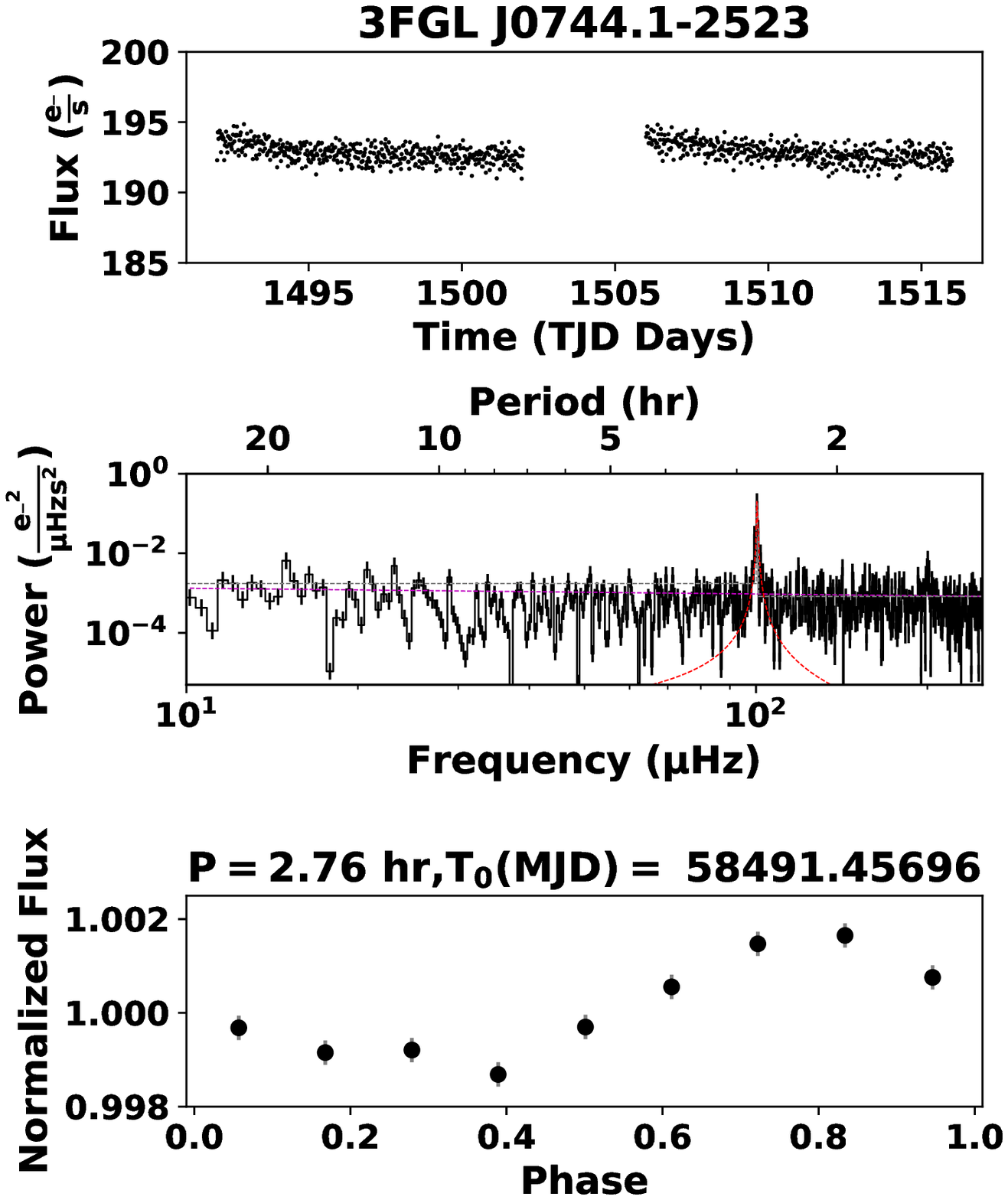}{0.48\textwidth}{(d):~Analysis result of \glseven}}
\caption{In all figures, top panel shows the unbinned \tess flux 
light curve; middle panel shows fitted PDS with {\it Bayesian} 
blocks and peak profiles; bottom panel shows the \tess light
 curves folded with periods shown above the panel.} 
\label{sources}
\end{figure*}

The \tess archival data are searched and obtained with 
\texttt{Astroquery}. \psroz, \glfive, \gltwo are observed under 
\texttt{TESS Guest Investigator Program}. For these three sources 
2\m cadence \texttt{timeseries} data are downloaded. For 
\glseven and \psrof, \texttt{TESSCUT} 30\m cadence FFI data within 
$3.85\arcmin \times 3.85\arcmin$ are downloaded. 
All data are analysed with \texttt{Lightkurve}\citep{lk18}.
Light curve files are generated 
with \texttt{to\_lightcurve()} from calibrated target-pixel files.
For GI proposal data {\it lightkurve}-defined aperture mask 
\texttt{pipeline\_mask} 
for GI proposal data. For \glseven \texttt{TESSCUT} data 
aperture mask is chosen manually depending upon the presence of 
peak profile in PDSs of the individual pixels.
The infinite or NaN values are excluded from light curves with 
\texttt{remove\_nans()}. The outliers above 3-$\sigma$ level in 
the light curves are clipped with \texttt{remove\_outliers()}. 
The threshold for \psroz is set to 6-$\sigma$ to retain the flaring events.
All cleaned unbinned light curves are plotted in the top panel of 
Fig.~\ref{sources}(a-d). In order to search periodicities in the optical 
flux Power Density Spectra are generated with \texttt{to\_periodogram} 
from cleaned unbinned light curves. Power Density Spectra are 
calculated using \texttt{Lomb-Scargle} method and normalized 
to \texttt{power spectral density} \citep{balona20}. Significant 
peak profiles in power density spectra are determined with 
{\it Bayesian block} analysis \citep{bayes} with 95\% statistical 
significance using \texttt{Astropy}. The peak profiles obtained 
from {\it Bayesian block} analysis are fitted with a {\it Lorentzian} 
profile \citep{belloni02} using \texttt{Scipy} to estimate the 
significance of the peak profiles. From curve fitting 
Q-value($\frac{\nu}{\Delta\nu}$) \citep{casella05}, 
RMS amplitude\footnote{\url{https://heasarc.gsfc.nasa.gov/docs/xte/recipes/pca\_fourier.html}} [see Eq.~\ref{eq_rms}] 
and reduced $\chi^2$ are calculated as goodness of fit. 
The 1-$\sigma$ error bars are estimated from covariance matrix.
The analysis results are shown in Table~\ref{table_ana}. 
In middle panel of Fig.~\ref{sources}(a-d) PDSs are shown in black color. 
The {\it Bayesian} blocks are shown in grey dashed lines. 
The {\it Lorentzian} peak profiles are shown in red, green, 
blue dashed lines. The continuum part of the PDS is fitted with 
power-law model and shown in magenta dashed line as an estimation 
of noise in the PDS.   
 
\begin{align}
RMS = & 100 \times \sqrt{\frac{A}{\overline{Flux}}} \%, \label{eq_rms} \\
where, ~ A = & \frac{\pi}{2} \times Normalization \times FWHM, \nonumber \\
 = & ~ Flux ~ under ~ Lorentzian ~ function; \nonumber \\
Normalization = & ~ Power ~ at ~ peak ~ frequency, \nonumber \\
FWHM = &~ Full ~ width ~ half ~ maxima. \nonumber  
\end{align}

\begin{figure*}[ht]
\centering
\gridline{
\fig{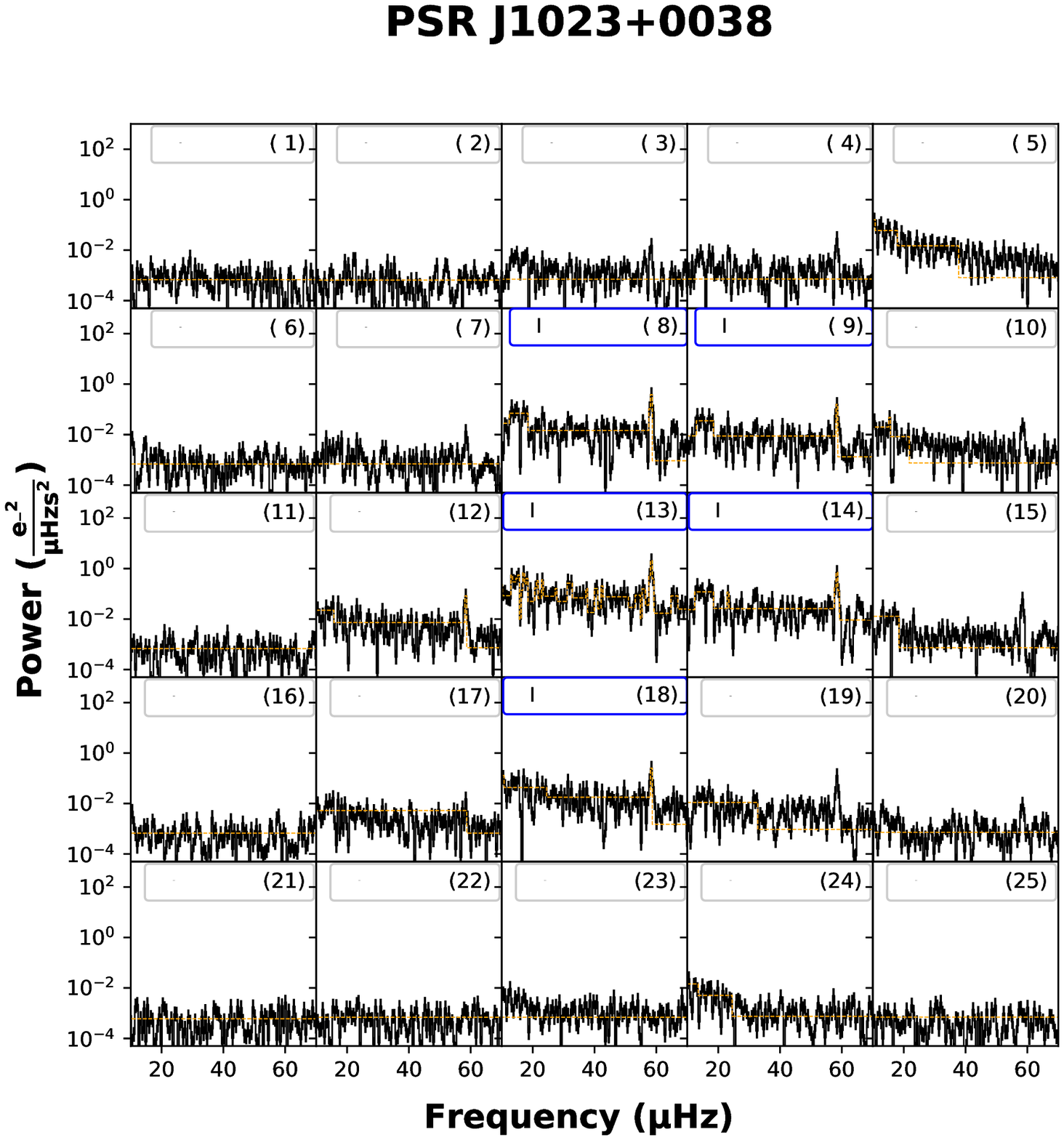}{0.48\textwidth}{(a):~Pixel-wise PDS of \psroz}
\fig{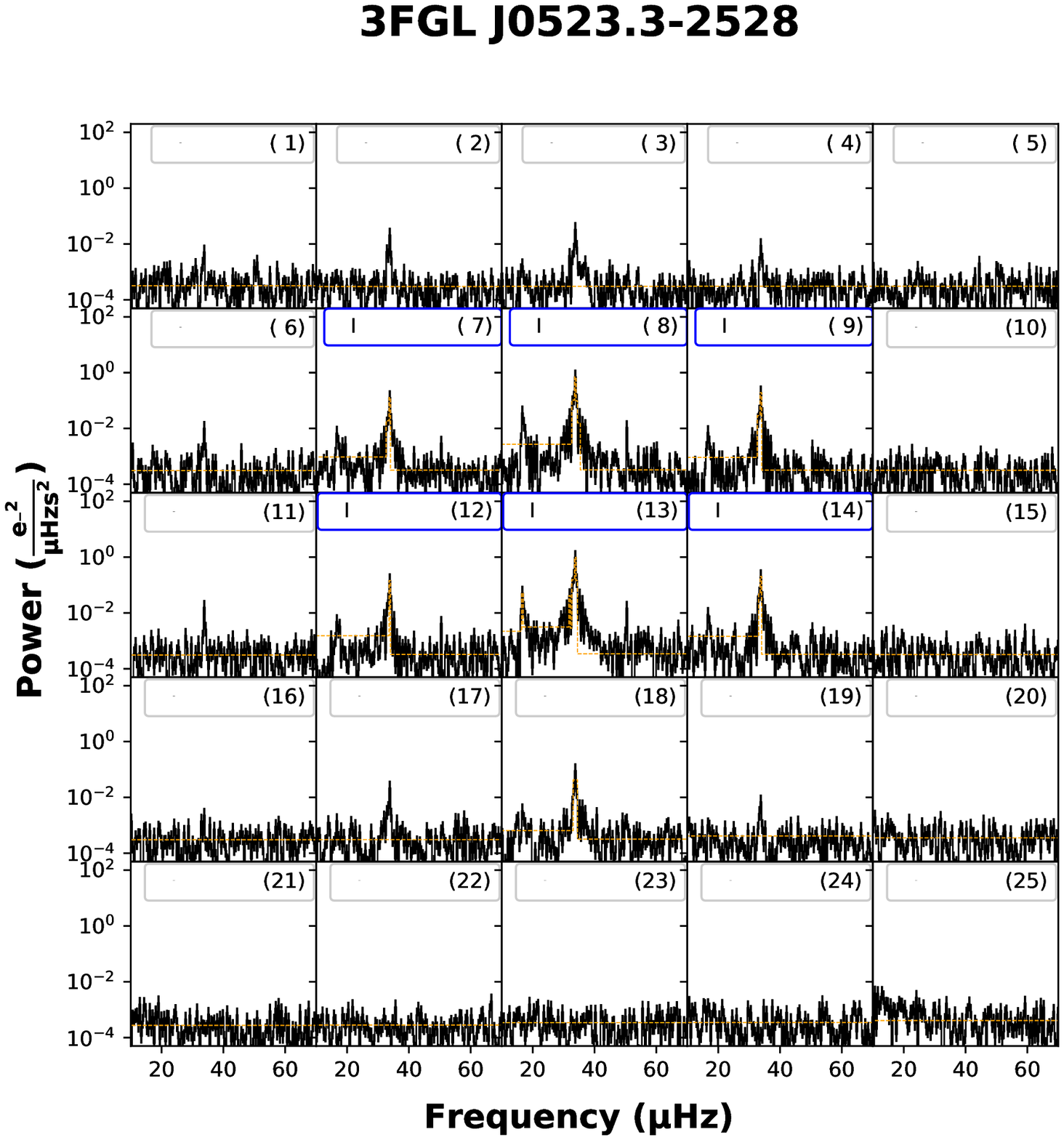}{0.48\textwidth}{(b):~Pixel-wise PDS of \glfive}}
\gridline{
\fig{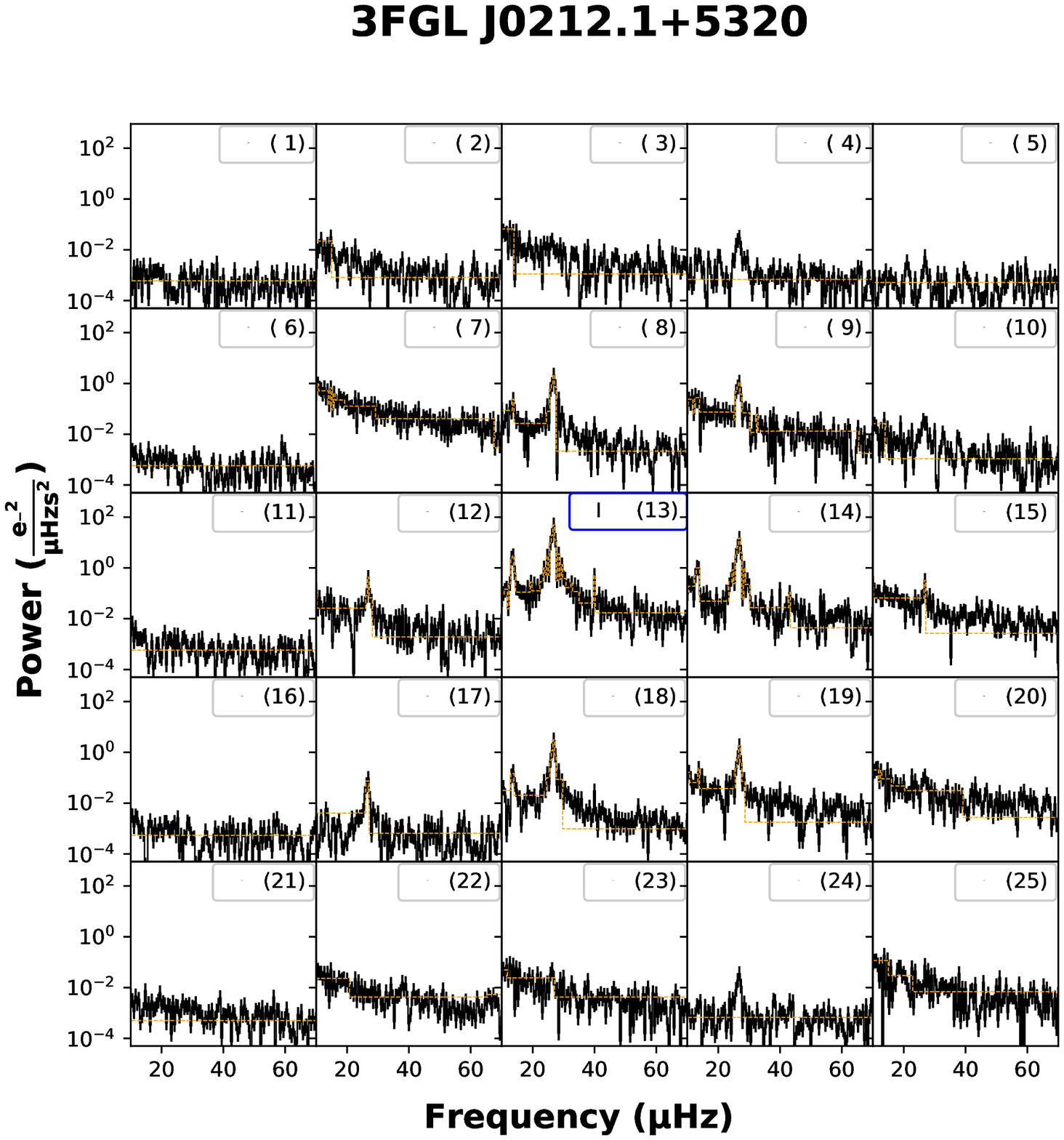}{0.48\textwidth}{(c):~Pixel-wise PDS of \gltwo}
\fig{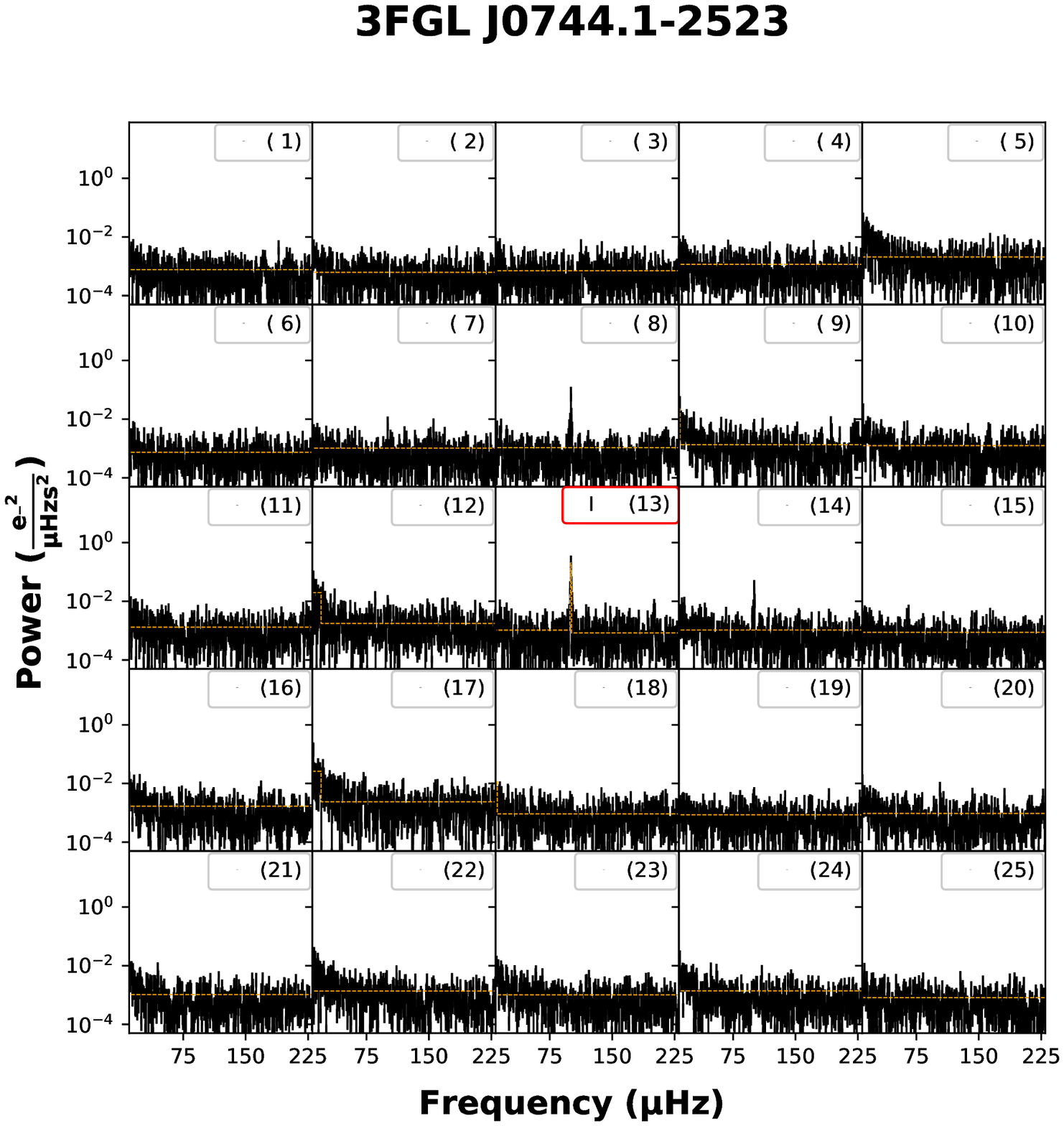}{0.48\textwidth}{(d):~Pixel-wise PDS of \glseven}}
\caption{Pixel-wise power density spectra (black) along with Bayesian 
blocks (orange) from the pixels around the aperture mask. Blue boxes 
represent pipeline defined aperture mask. Red box represent 
manually chosen aperture mask with strongest peak profile.}
\label{pixel}
\end{figure*}

Before detailed analysis for individual sources, one would 
like to first verify that the obtained periodicity in the 
PDS indeed comes from the corresponding redback positions. 
Hence, PDSs of pixels around aperture mask 
(i.e., neighborhood pixels) are plotted with Bayesian blocks, 
as shown in Fig.~\ref{pixel}(a-d). 
In Fig.~\ref{pixel}(a-c) the blue boxes 
mark the pixels included in the pipeline-defined aperture mask. 
For \glseven in Fig.~\ref{pixel}(d), the $5 \times 5$ pixel PDS plot 
with peak profiles at the same frequencies are observed. 
The pixel with the strongest peak profile is taken as aperture mask for the 
further analysis of \glseven \texttt{TESSCUT} data (marked with a red box).
For \psrof in Fig.~\ref{j1417}(a), the $5 \times 5$ pixel 
PDS plot is shown where no peak profile is observed. 
The pixel coordinates are projected on sky coordinates and 
\psrof sky coordinates coincides with pixel \#13, which is 
marked with a blue box.
This pixel is taken as aperture mask for further analysis of 
\psrof \texttt{TESSCUT} data.
  
With the obtained orbit-related periodicities of redback systems 
(candidates) using \tess data, we then fold the \tess light curves 
for four sources with the orbital period, 
as shown in bottom panel of Fig.~\ref{sources}(a-d) 
and Fig.~\ref{j1417}(b). The low frequency trends are removed 
from data using \texttt{Savitzky-Golay filter}.
The flux is normalized to the median flux of each source. 
The phase zero ($T_\mathrm{0}$) is 
set to the inferior conjunction (when the companion is between the 
pulsar and the observer), following~\citet{ta05} for \psroz, 
\citet{linares17} for \gltwo, \citet{strader14} for \glfive and  
\citet{salvetti17} for \glseven.

\section{Results}

\begin{deluxetable*}{cccccc}
\tablenum{2}
\tablecaption{PDS fitting results: (1) Source names, (2) Orbital period calculated from the peak frequency of {\it Lorentzian} 
profile, (3) Full Width Half Maximum of {\it Lorentzian}, (4) Q-value, (5) RMS amplitude (\%), (6) Reduced $\chi^2$ value.   
\label{table_ana}}
\tablewidth{0pt}
\tablehead{
\colhead{Source Name} & \colhead{Period} & \colhead{FWHM} & \colhead{Q-value} & \colhead{RMS} &  \colhead{$\chi^2_\nu$} \\
\colhead{} & \colhead{(hr)} & \colhead{(hr)} & \colhead{} & \colhead{(\%)} &  \colhead{(dof)} 
}
\decimalcolnumbers
\startdata
PSR J1023+0038 & $4.7816 \pm 0.0015$ & $0.0274 \pm 0.0027$ & 174.57 & 19.22 & 2.17(40) \\
\hline                                     
{               } & $8.2768 \pm 0.0020$ & $0.0207 \pm 0.0010$ & 400.62 & 8.99 & 0.65(55) \\
3FGL J0523.3-2528 & $16.4411 \pm 0.0149$ & $0.0994 \pm 0.0068$ & 165.42 & 2.68 & 0.12(19) \\
{               } & $5.5028 \pm 0.0037$ & $0.0251 \pm 0.0041$ & 219.18 & 3.02 & 0.07(42) \\
\hline                                     
3FGL J0212.1+5320 & $10.4659 \pm 0.0051$ & $0.0167 \pm 0.0751$ & 625.48 & 15.74 & 0.08(18) \\
\hline                                   
3FGL J0744.1-2523 & $2.7634 \pm 0.0008$ & $0.0034 \pm 0.0015$ & 815.86 & 0.70 & 0.01(38) \\
\hline                                     
\enddata
\end{deluxetable*}

\subsection{\psroz}
In \tess data \psroz shows a median flux around 60\eps for 
around 26 days. The {\it Lomb-Scargle} periodogram and its 
{\it Bayesian} analysis shows a peak between 50-60\muhz with 
power around 10\mupo~(see Fig.~\ref{sources}(a)). 
From these peak profile parameters a 
period of 4.7816$\pm$0.0015\hr (i.e., corresponding to the 
reported orbital period) 
is obtained. The Q-value for this peak profile is 174.57 with 
19.22\% of flux photons responsible for the orbital modulation. 
The goodness of fit is 2.17(40). No other signature of periodicity 
is observed in the PDS. The continuum part before and after the 
orbital period is fitted with power-law models. The power-law index of 
lower-frequency part is $\Gamma_1 = 0.65\pm0.14$ and upper-frequency 
part is $\Gamma_2 = 1.2\pm0.05$. 

Short-time flares are also seen, and are occurring as often 
as that reported in \citet{kennedy18, papitto19}.

\subsection{\glfive}
Fig.~\ref{sources}(b) shows the light curve of \glfive. 
\tess has observed \glfive for a 
total of 50 days in two consecutive observation sectors (\#5 and \#6). 
We checked the full-frame images, finding that different 
pipeline-defined aperture masks are used during two sectors.
To maintain consistency, we here use the aperture mask defined 
by the pipeline for sector~\#5 for both sectors.
The {\it Lomb-Scargle} periodogram and its {\it Bayesian} analysis 
shows 3 peak profiles. The fundamental frequency,
shown in red dashed line, is observed at 33.56\muhz which 
represents a period of 8.28\hr with Q-value of 400.62 
and 8.99\% of observed flux photons responsible for the 
orbital motion. Around 16.89\muhz, blue dashed line, half-harmonic 
is observed which yields an oscillation of 16.44\hr with 
Q-value 165.42 and 2.68\% observed flux involved in it. 
The goodness of fit is 0.12(19). The first harmonic of this 
orbital motion is also observed in the PDS at 67.26\muhz. 
This peak is less than 95\% statistical significant. 
This peak is ignored. Another peak profile is observed at 50.48\muhz 
shown in green dashed line. This peak profile represent 
another orbital motion for 5.5\hr with 219.18 Q-value 
and 3.02\% RMS amplitude.
The continuum at lower frequency shows power-law index 
of $\Gamma_1 = 0.02\pm0.9$ and upper-frequency shows 
$\Gamma_2 = 0.21\pm0.03$. 

 \tess data reveal the orbital period~\citep[16.5~hr,][]{strader14} and 
also a period 
of 8.27\hr (one half of the orbital period) which is characteristic 
of ellipsoidal variation (EV).
In the \tess PDS another periodicity of 5.5\hr is 
obtained with significant Q-value and RMS. This periodicity may 
represent some periodicity in the \glfive system which has not 
been reported before. Around 4.14\hr(i.e.~67\muhz) another 
peak may be seen which 
is likely a harmonic of the EV frequency. This peak profile is 
below 95\% {\it Bayesian} block significance. 

\subsection{\gltwo}
Fig.~\ref{sources}(c) shows the light curve of \gltwo. 
This light curve is affected by background 
scattering\footnote{\url{https://archive.stsci.edu/missions/tess/doc/tess\_drn/tess\_sector\_18\_drn25\_v01.pdf}} and so we select those part of 
the data without severe scattering. Between TJD 1790-1795 a flare 
up to 300\eps is observed, after that the flux increases gradually. 
We believe such an increase is genuine as we also observe changes in 
the phase profile as the brightness changes (but not during the 
`steady flux' state seen after TJD 1803).
{\it Bayesian} block analysis of the PDS shows a peak profile
 around 26.53\muhz. This represents a period of 10.47\hr
with 625.48 Q-value and 15.74\% flux contribution for the orbital motion. 
The continuum at upper-frequency shows power-law index of 
$\Gamma = 2.34\pm0.10$.

This value is about twice that of the orbital period of 
0.87 days, or 20.88\hr \citep{li16, linares17}. 
In the \tess PDS pulse profile, no peak is located around 
20.88\hr/13.3\muhz. However, a break in frequency domain is 
observed around this frequency which may cover a possible periodicity here.

\subsection{\glseven}
In Fig.~\ref{sources}(d) the analysis 
result of \glseven Sector. \#7 \texttt{TESSCUT} data is shown. 
The flux rate is within 190 \eps and 195\eps 
for this observation. No strong flares are observed. 
Here in the PDS a pulse profile is observed from 
{\it Bayesian} block analysis. 
At 100.52\muhz one peak profile, is observed, shown in red color. 
This peak represents 2.76\hr period with Q-value = 815.86 and RMS = 0.7\%.
The flat continuum of PDS shows a power-law index of $\Gamma = 0.15\pm0.14$.  

\subsection{\psrof}    
\label{result_1417}
\begin{figure*}[ht]
\centering
\gridline{
\fig{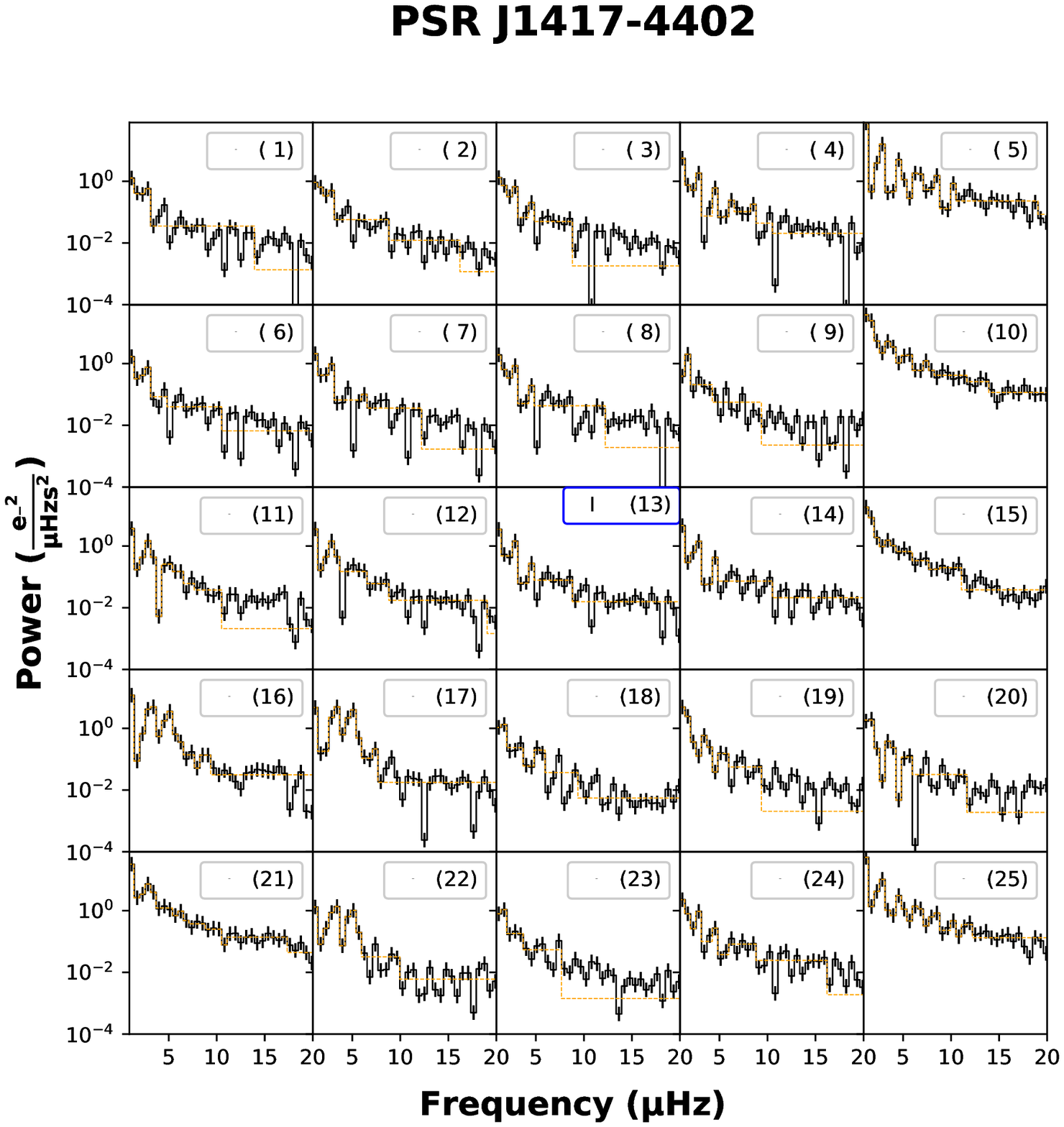}{0.51\textwidth}{(a):~Pixel-wise PDS of \psrof}
\fig{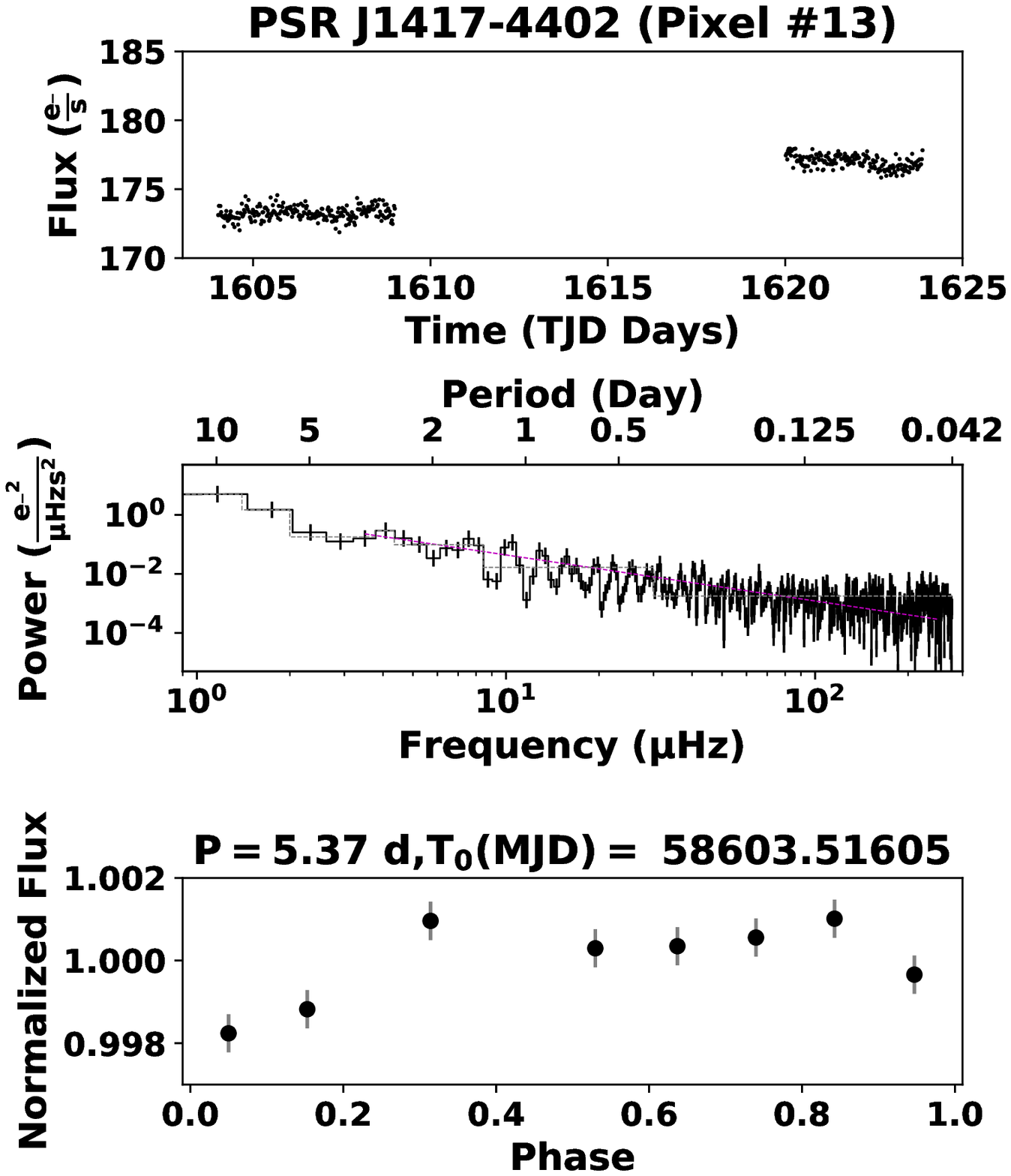}{0.45\textwidth}{(b):~Analysis result of \psrof}}
\caption{Pixel-wise Power Density Spectra (black) along with Bayesian 
blocks(orange), around the sky region of \psrof. The sky location of 
\psrof superimposes on Pixel \#13 (marked with blue). 
\texttt{TESSCUT} data from Pixel \#13 are further analysed.}
\label{j1417}
\end{figure*}

In Fig.~\ref{j1417}(a) pixel-wise PDS diagram for  
\psrof is shown where no significant periodicity can be seen.
The pixel \#13 is analysed and shown in Fig.~\ref{j1417}(b). 
We folded the light curve with the reported
orbital period 5.37372(3) day \citep{camilo16}, corresponding 
to a frequency 2.15\muhz, as shown in Fig.~\ref{j1417}(c). 
The uneven binning in the folded lightcurve is due to the time 
gap in the light curve.

\section{Discussion}
The orbital period \psroz obtained in this work is formally 
not the same as reported earlier. A caveat in this work is that 
we did not exclude the flares from the original light curve, 
and so the flaring signals may contaminate the periodic signal. 
Since this work is mainly to demonstrate the TESS capability 
to search for periodicities from a large sample of sources, 
we did not attempt to remove the flares, as has been done in 
\citet{kennedy18, papitto19} for the K2 80-day 
light curve with 58.8\s cadence.

Two peaks resulting from the secondary star's ellipsoidal 
modulation can be seen in the orbital phased light curves 
of \glfive and \gltwo (see, bottom panel of Fig.~\ref{sources}(b-c)). 
Other than the two maxima and minima consistent with 
previous observations \citep{linares17, li16, strader14}, 
we find asymmetries in both minima and maxima. It confirms the 
findings of \citet{li16} and \citet{linares17} for \gltwo and 
the asymmetry is 
first discovered for \glfive \citep[whereas in previously reported 
light curve no significant asymmetry can be seen between the 
two maxima due to lower photometric accuracy][]{strader14}. 
While the unequal minima might be partly explained by limb- and 
gravity-darkening effects, the unequal maxima distinguishes these 
two pulsar binaries from normal ellipsoidal variables. Various models 
based on additional light source have been proposed to explain this 
phenomenon, such as off-center heating from an intra-binary 
shock \citep{rs16} or star-spot activity \citep{va16}.

For \gltwo, one may compare the EV-derived orbital period seen 
in \tess data (20.9318$\pm$0.0102~hrs) taken around the mean MJD 58802) 
with that reported earlier in \citet{li16} (20.8698(1)~hrs taken at 
a mean MJD 57357) and \citet{linares17} (20.8692(36)~hrs, taken at
 a mean MJD 57218), showing plausible hint of increasing orbital 
period over time, but caution must be taken to over-interpret 
such a possibility.

For \glseven, an optical period of $2.7701 \pm 0.0012$~hours is 
reported in \citet{salvetti17}. The periodicity value 
$2.76342 \pm 0.00079$ obtained from our analysis 
is comparable with the literature value. 

In the case of \psrof there is a flux gap within the light curve. Here we focus on periodicity analysis.
The orbital period of \psrof is reported to be 
around 5.37372(3) days \citep{camilo16}.
The effective exposure of the data analysed is around 10 days, a mere two orbital cycles. Combined with the flux gap and contamination from a nearby bright star, it may explain the non-detection of the period.

\section{Conclusions}
In this paper, \tess data from five pulsar binaries 
(or candidates thereof), are analysed. 
In two cases (i.e. \psroz and \glseven) the period can be identified with 
the orbital period itself where the optical brightness variation 
may be due to pulsar irradiation. For other two cases 
(i.e. \glfive, \gltwo), frequency peaks 
corresponding to literature values can be obtained and optical 
modulations are observed at half the orbital period values,
 which are EV signatures.

In summary, \tess data revealed periodicities seen 
in four previously reported redback-like systems. 
Analysis of \tess data for other similar binary 
systems will explore more information.

\vspace{5mm}
\acknowledgments
PSP acknowledges SYSU-Postdoctoral Fellowship. PSP and PHT are supported 
by NSFC through grants 11633007, 11661161010 and U1731136. KLL is supported by the Ministry of Science and Technology (MOST) of the Republic of China (Taiwan) through grant 108-2112-M-007-025-MY3. CYH is supported by the National Research Foundation of Korea through grant 2016R1A5A1013277 and 2019R1F1A1062071. AKHK is supported by MOST of the Republic of China (Taiwan) through grant 105-2119-M-007-028-MY3.
We thank Science Processing Operations Center (SPOC-NASA) 
and Mikulski Archive for Space Telescopes (MAST-STScI) 
for archival of the \tess data.

\vspace{5mm}
\facilities{{\em TESS}\citep{tess15}}

\software{
Astropy\citep[v3.2.1][]{astropy13}, Astroquery\cite[v0.3.10]{astroquery13}, SciPy\citep[v1.2.1][]{scipy19}, 
Lightkurve\citep[v1.6][]{lk18}, NumPy\citep[v1.17.3][]{numpy11}.
}

\vskip 15cm
\bibliography{tess_ref}{}

\begin{thebibliography}{}
\expandafter\ifx\csname natexlab\endcsname\relax\def\natexlab#1{#1}\fi
\providecommand{\url}[1]{\href{#1}{#1}}
\providecommand{\dodoi}[1]{doi:~\href{http://doi.org/#1}{\nolinkurl{#1}}}
\providecommand{\doeprint}[1]{\href{http://ascl.net/#1}{\nolinkurl{http://ascl.net/#1}}}
\providecommand{\doarXiv}[1]{\href{https://arxiv.org/abs/#1}{\nolinkurl{https://arxiv.org/abs/#1}}}

\bibitem[{{Abdo} {et~al.}(2010){Abdo}, {Ackermann}, {Ajello}, {Allafort},
  {Antolini}, {Atwood}, {Axelsson}, {Baldini}, {Ballet}, {Barbiellini},
  {Bastieri}, {Baughman}, {Bechtol}, {Bellazzini}, {Belli}, {Berenji},
  {Bisello}, {Blandford}, {Bloom}, {Bonamente}, {Bonnell}, {Borgland},
  {Bouvier}, {Bregeon}, {Brez}, {Brigida}, {Bruel}, {Burnett}, {Busetto},
  {Buson}, {Caliand ro}, {Cameron}, {Campana}, {Canadas}, {Caraveo},
  {Carrigan}, {Casandjian}, {Cavazzuti}, {Ceccanti}, {Cecchi}, {{\c{C}}elik},
  {Charles}, {Chekhtman}, {Cheung}, {Chiang}, {Cillis}, {Ciprini}, {Claus},
  {Cohen-Tanugi}, {Conrad}, {Corbet}, {Davis}, {DeKlotz}, {den Hartog},
  {Dermer}, {de Angelis}, {de Luca}, {de Palma}, {Digel}, {Dormody}, {Silva},
  {Drell}, {Dubois}, {Dumora}, {Fabiani}, {Farnier}, {Favuzzi}, {Fegan},
  {Ferrara}, {Focke}, {Fortin}, {Frailis}, {Fukazawa}, {Funk}, {Fusco},
  {Gargano}, {Gasparrini}, {Gehrels}, {Germani}, {Giavitto}, {Giebels},
  {Giglietto}, {Giommi}, {Giordano}, {Giroletti}, {Glanzman}, {Godfrey},
  {Grenier}, {Grondin}, {Grove}, {Guillemot}, {Guiriec}, {Gustafsson},
  {Hadasch}, {Hanabata}, {Harding}, {Hayashida}, {Hays}, {Healey}, {Hill},
  {Horan}, {Hughes}, {Iafrate}, {J{\'o}hannesson}, {Johnson}, {Johnson},
  {Johnson}, {Johnson}, {Kamae}, {Katagiri}, {Kataoka}, {Kawai}, {Kerr},
  {Kn{\"o}dlseder}, {Kocevski}, {Kuss}, {Lande}, {Landriu}, {Latronico}, {Lee},
  {Lemoine-Goumard}, {Lionetto}, {Llena Garde}, {Longo}, {Loparco}, {Lott},
  {Lovellette}, {Lubrano}, {Madejski}, {Makeev}, {Marangelli}, {Marelli},
  {Massaro}, {Mazziotta}, {McConville}, {McEnery}, {Michelson}, {Minuti},
  {Mitthumsiri}, {Mizuno}, {Moiseev}, {Mongelli}, {Monte}, {Monzani},
  {Moretti}, {Morselli}, {Moskalenko}, {Murgia}, {Nakajima}, {Nakamori},
  {Naumann-Godo}, {Nolan}, {Norris}, {Nuss}, {Ohno}, {Ohsugi}, {Omodei},
  {Orlando}, {Ormes}, {Ozaki}, {Paccagnella}, {Paneque}, {Panetta}, {Parent},
  {Pelassa}, {Pepe}, {Pesce-Rollins}, {Pinchera}, {Piron}, {Porter}, {Poupard},
  {Rain{\`o}}, {Rando}, {Ray}, {Razzano}, {Razzaque}, {Rea}, {Reimer},
  {Reimer}, {Reposeur}, {Ripken}, {Ritz}, {Rochester}, {Rodriguez}, {Romani},
  {Roth}, {Sadrozinski}, {Salvetti}, {Sanchez}, {Sand er}, {Saz Parkinson},
  {Scargle}, {Schalk}, {Scolieri}, {Sgr{\`o}}, {Shaw}, {Siskind}, {Smith},
  {Smith}, {Spandre}, {Spinelli}, {Starck}, {Stephens}, {Striani}, {Strickman},
  {Strong}, {Suson}, {Tajima}, {Takahashi}, {Takahashi}, {Tanaka}, {Thayer},
  {Thayer}, {Thompson}, {Tibaldo}, {Tibolla}, {Tinebra}, {Torres}, {Tosti},
  {Tramacere}, {Uchiyama}, {Usher}, {Van Etten}, {Vasileiou}, {Vilchez},
  {Vitale}, {Waite}, {Wallace}, {Wang}, {Watters}, {Winer}, {Wood}, {Yang},
  {Ylinen}, {Ziegler}, \& {Fermi LAT Collaboration}}]{abdo10}
{Abdo}, A.~A., {Ackermann}, M., {Ajello}, M., {et~al.} 2010, \apjs, 188, 405,
  \dodoi{10.1088/0067-0049/188/2/405}

\bibitem[{{Acero} {et~al.}(2015){Acero}, {Ackermann}, {Ajello}, {Albert},
  {Atwood}, {Axelsson}, {Baldini}, {Ballet}, {Barbiellini}, {Bastieri},
  {Belfiore}, {Bellazzini}, {Bissaldi}, {Blandford}, {Bloom}, {Bogart},
  {Bonino}, {Bottacini}, {Bregeon}, {Britto}, {Bruel}, {Buehler}, {Burnett},
  {Buson}, {Caliand ro}, {Cameron}, {Caputo}, {Caragiulo}, {Caraveo},
  {Casandjian}, {Cavazzuti}, {Charles}, {Chaves}, {Chekhtman}, {Cheung},
  {Chiang}, {Chiaro}, {Ciprini}, {Claus}, {Cohen-Tanugi}, {Cominsky}, {Conrad},
  {Cutini}, {D'Ammando}, {de Angelis}, {DeKlotz}, {de Palma}, {Desiante},
  {Digel}, {Di Venere}, {Drell}, {Dubois}, {Dumora}, {Favuzzi}, {Fegan},
  {Ferrara}, {Finke}, {Franckowiak}, {Fukazawa}, {Funk}, {Fusco}, {Gargano},
  {Gasparrini}, {Giebels}, {Giglietto}, {Giommi}, {Giordano}, {Giroletti},
  {Glanzman}, {Godfrey}, {Grenier}, {Grondin}, {Grove}, {Guillemot}, {Guiriec},
  {Hadasch}, {Harding}, {Hays}, {Hewitt}, {Hill}, {Horan}, {Iafrate}, {Jogler},
  {J{\'o}hannesson}, {Johnson}, {Johnson}, {Johnson}, {Johnson}, {Kamae},
  {Kataoka}, {Katsuta}, {Kuss}, {La Mura}, {Land riu}, {Larsson}, {Latronico},
  {Lemoine-Goumard}, {Li}, {Li}, {Longo}, {Loparco}, {Lott}, {Lovellette},
  {Lubrano}, {Madejski}, {Massaro}, {Mayer}, {Mazziotta}, {McEnery},
  {Michelson}, {Mirabal}, {Mizuno}, {Moiseev}, {Mongelli}, {Monzani},
  {Morselli}, {Moskalenko}, {Murgia}, {Nuss}, {Ohno}, {Ohsugi}, {Omodei},
  {Orienti}, {Orlando}, {Ormes}, {Paneque}, {Panetta}, {Perkins},
  {Pesce-Rollins}, {Piron}, {Pivato}, {Porter}, {Racusin}, {Rando}, {Razzano},
  {Razzaque}, {Reimer}, {Reimer}, {Reposeur}, {Rochester}, {Romani},
  {Salvetti}, {S{\'a}nchez-Conde}, {Saz Parkinson}, {Schulz}, {Siskind},
  {Smith}, {Spada}, {Spandre}, {Spinelli}, {Stephens}, {Strong}, {Suson},
  {Takahashi}, {Takahashi}, {Tanaka}, {Thayer}, {Thayer}, {Thompson},
  {Tibaldo}, {Tibolla}, {Torres}, {Torresi}, {Tosti}, {Troja}, {Van Klaveren},
  {Vianello}, {Winer}, {Wood}, {Wood}, {Zimmer}, \& {Fermi-LAT
  Collaboration}}]{acero15}
{Acero}, F., {Ackermann}, M., {Ajello}, M., {et~al.} 2015, \apjs, 218, 23,
  \dodoi{10.1088/0067-0049/218/2/23}

\bibitem[{{Archibald} {et~al.}(2009){Archibald}, {Stairs}, {Ransom}, {Kaspi},
  {Kondratiev}, {Lorimer}, {McLaughlin}, {Boyles}, {Hessels}, {Lynch}, {van
  Leeuwen}, {Roberts}, {Jenet}, {Champion}, {Rosen}, {Barlow}, {Dunlap}, \&
  {Remillard}}]{arch09}
{Archibald}, A.~M., {Stairs}, I.~H., {Ransom}, S.~M., {et~al.} 2009, Science,
  324, 1411, \dodoi{10.1126/science.1172740}

\bibitem[{{Astropy Collaboration} {et~al.}(2013){Astropy Collaboration},
  {Robitaille}, {Tollerud}, {Greenfield}, {Droettboom}, {Bray}, {Aldcroft},
  {Davis}, {Ginsburg}, {Price-Whelan}, {Kerzendorf}, {Conley}, {Crighton},
  {Barbary}, {Muna}, {Ferguson}, {Grollier}, {Parikh}, {Nair}, {Unther},
  {Deil}, {Woillez}, {Conseil}, {Kramer}, {Turner}, {Singer}, {Fox}, {Weaver},
  {Zabalza}, {Edwards}, {Azalee Bostroem}, {Burke}, {Casey}, {Crawford},
  {Dencheva}, {Ely}, {Jenness}, {Labrie}, {Lim}, {Pierfederici}, {Pontzen},
  {Ptak}, {Refsdal}, {Servillat}, \& {Streicher}}]{astropy13}
{Astropy Collaboration}, {Robitaille}, T.~P., {Tollerud}, E.~J., {et~al.} 2013,
  \aap, 558, A33, \dodoi{10.1051/0004-6361/201322068}

\bibitem[{{Balona} \& {Ozuyar}(2020)}]{balona20}
{Balona}, L.~A., \& {Ozuyar}, D. 2020, \mnras, 493, 2528,
  \dodoi{10.1093/mnras/staa389}

\bibitem[{{Bassa} {et~al.}(2014){Bassa}, {Patruno}, {Hessels}, {Keane},
  {Monard}, {Mahony}, {Bogdanov}, {Corbel}, {Edwards}, {Archibald}, {Janssen},
  {Stappers}, \& {Tendulkar}}]{bassa14}
{Bassa}, C.~G., {Patruno}, A., {Hessels}, J.~W.~T., {et~al.} 2014, \mnras, 441,
  1825, \dodoi{10.1093/mnras/stu708}

\bibitem[{{Belloni} {et~al.}(2002){Belloni}, {Psaltis}, \& {van der
  Klis}}]{belloni02}
{Belloni}, T., {Psaltis}, D., \& {van der Klis}, M. 2002, \apj, 572, 392,
  \dodoi{10.1086/340290}

\bibitem[{{Camilo} {et~al.}(2016){Camilo}, {Reynolds}, {Ransom}, {Halpern},
  {Bogdanov}, {Kerr}, {Ray}, {Cordes}, {Sarkissian}, {Barr}, \&
  {Ferrara}}]{camilo16}
{Camilo}, F., {Reynolds}, J.~E., {Ransom}, S.~M., {et~al.} 2016, \apj, 820, 6,
  \dodoi{10.3847/0004-637X/820/1/6}

\bibitem[{{Casella} {et~al.}(2005){Casella}, {Belloni}, \&
  {Stella}}]{casella05}
{Casella}, P., {Belloni}, T., \& {Stella}, L. 2005, \apj, 629, 403,
  \dodoi{10.1086/431174}

\bibitem[{{De Vito} {et~al.}(2019){De Vito}, {Horvath}, \&
  {Benvenuto}}]{devito19}
{De Vito}, M.~A., {Horvath}, J.~E., \& {Benvenuto}, O.~G. 2019, \mnras, 483,
  4495, \dodoi{10.1093/mnras/sty3476}

\bibitem[{{Deller} {et~al.}(2012){Deller}, {Archibald}, {Brisken},
  {Chatterjee}, {Janssen}, {Kaspi}, {Lorimer}, {Lyne}, {McLaughlin}, {Ransom},
  {Stairs}, \& {Stappers}}]{deller12}
{Deller}, A.~T., {Archibald}, A.~M., {Brisken}, W.~F., {et~al.} 2012, \apjl,
  756, L25, \dodoi{10.1088/2041-8205/756/2/L25}

\bibitem[{{Dorn-Wallenstein} {et~al.}(2019){Dorn-Wallenstein}, {Levesque}, \&
  {Davenport}}]{dorn19}
{Dorn-Wallenstein}, T.~Z., {Levesque}, E.~M., \& {Davenport}, J. R.~A. 2019,
  \apj, 878, 155, \dodoi{10.3847/1538-4357/ab223f}

\bibitem[{Ginsburg {et~al.}(2013)Ginsburg, Robitaille, Parikh, Deil, Mirocha,
  Woillez, Svoboda, Willett, Allen, Grollier, Persson, \& Shiga}]{astroquery13}
Ginsburg, A., Robitaille, T., Parikh, M., {et~al.} 2013,
  \dodoi{10.6084/m9.figshare.805208.v2}

\bibitem[{{Guillemot} {et~al.}(2012){Guillemot}, {Freire}, {Cognard},
  {Johnson}, {Takahashi}, {Kataoka}, {Desvignes}, {Camilo}, {Ferrara},
  {Harding}, {Janssen}, {Keith}, {Kerr}, {Kramer}, {Parent}, {Ransom}, {Ray},
  {Saz Parkinson}, {Smith}, {Stappers}, \& {Theureau}}]{guil12}
{Guillemot}, L., {Freire}, P.~C.~C., {Cognard}, I., {et~al.} 2012, \mnras, 422,
  1294, \dodoi{10.1111/j.1365-2966.2012.20694.x}

\bibitem[{{Hui} \& {Li}(2019)}]{Hui19}
{Hui}, C.~Y., \& {Li}, K.~L. 2019, Galaxies, 7, 93,
  \dodoi{10.3390/galaxies7040093}

\bibitem[{{Kennedy} {et~al.}(2018){Kennedy}, {Clark}, {Voisin}, \&
  {Breton}}]{kennedy18}
{Kennedy}, M.~R., {Clark}, C.~J., {Voisin}, G., \& {Breton}, R.~P. 2018,
  \mnras, 477, 1120, \dodoi{10.1093/mnras/sty731}

\bibitem[{{Li} {et~al.}(2016){Li}, {Kong}, {Hou}, {Mao}, {Strader}, {Chomiuk},
  \& {Tremou}}]{li16}
{Li}, K.-L., {Kong}, A. K.~H., {Hou}, X., {et~al.} 2016, \apj, 833, 143,
  \dodoi{10.3847/1538-4357/833/2/143}

\bibitem[{{Lightkurve Collaboration} {et~al.}(2018){Lightkurve Collaboration},
  {Cardoso}, {Hedges}, {Gully-Santiago}, {Saunders}, {Cody}, {Barclay}, {Hall},
  {Sagear}, {Turtelboom}, {Zhang}, {Tzanidakis}, {Mighell}, {Coughlin}, {Bell},
  {Berta-Thompson}, {Williams}, {Dotson}, \& {Barentsen}}]{lk18}
{Lightkurve Collaboration}, {Cardoso}, J.~V.~d.~M., {Hedges}, C., {et~al.}
  2018, {Lightkurve: Kepler and TESS time series analysis in Python},
  Astrophysics Source Code Library.
\newblock \doeprint{1812.013}

\bibitem[{{Linares}(2018)}]{linares18}
{Linares}, M. 2018, \mnras, 473, L50, \dodoi{10.1093/mnrasl/slx153}

\bibitem[{{Linares} {et~al.}(2017){Linares}, {Miles-P{\'a}ez},
  {Rodr{\'\i}guez-Gil}, {Shahbaz}, {Casares}, {Fari{\~n}a}, \&
  {Karjalainen}}]{linares17}
{Linares}, M., {Miles-P{\'a}ez}, P., {Rodr{\'\i}guez-Gil}, P., {et~al.} 2017,
  \mnras, 465, 4602, \dodoi{10.1093/mnras/stw3057}

\bibitem[{{McConnell} {et~al.}(2015){McConnell}, {Callanan}, {Kennedy},
  {Hurley}, {Garnavich}, \& {Menzies}}]{mcconnell15}
{McConnell}, O., {Callanan}, P.~J., {Kennedy}, M., {et~al.} 2015, \mnras, 451,
  3468, \dodoi{10.1093/mnras/stv1197}

\bibitem[{{Papitto} {et~al.}(2013){Papitto}, {Ferrigno}, {Bozzo}, {Rea},
  {Pavan}, {Burderi}, {Burgay}, {Campana}, {di Salvo}, {Falanga},
  {Filipovi{\'c}}, {Freire}, {Hessels}, {Possenti}, {Ransom}, {Riggio},
  {Romano}, {Sarkissian}, {Stairs}, {Stella}, {Torres}, {Wieringa}, \&
  {Wong}}]{papitto13}
{Papitto}, A., {Ferrigno}, C., {Bozzo}, E., {et~al.} 2013, \nat, 501, 517,
  \dodoi{10.1038/nature12470}

\bibitem[{{Papitto} {et~al.}(2019){Papitto}, {Ambrosino}, {Stella}, {Torres},
  {Coti Zelati}, {Ghedina}, {Meddi}, {Sanna}, {Casella}, {Dallilar},
  {Eikenberry}, {Israel}, {Onori}, {Piranomonte}, {Bozzo}, {Burderi},
  {Campana}, {de Martino}, {Di Salvo}, {Ferrigno}, {Rea}, {Riggio}, {Serrano},
  {Veledina}, \& {Zampieri}}]{papitto19}
{Papitto}, A., {Ambrosino}, F., {Stella}, L., {et~al.} 2019, \apj, 882, 104,
  \dodoi{10.3847/1538-4357/ab2fdf}

\bibitem[{{Patruno} {et~al.}(2014){Patruno}, {Archibald}, {Hessels},
  {Bogdanov}, {Stappers}, {Bassa}, {Janssen}, {Kaspi}, {Tendulkar}, \&
  {Lyne}}]{patruno14}
{Patruno}, A., {Archibald}, A.~M., {Hessels}, J.~W.~T., {et~al.} 2014, \apjl,
  781, L3, \dodoi{10.1088/2041-8205/781/1/L3}

\bibitem[{{Petrov} {et~al.}(2013){Petrov}, {Mahony}, {Edwards}, {Sadler},
  {Schinzel}, \& {McConnell}}]{petrov13}
{Petrov}, L., {Mahony}, E.~K., {Edwards}, P.~G., {et~al.} 2013, \mnras, 432,
  1294, \dodoi{10.1093/mnras/stt550}

\bibitem[{{Ricker} {et~al.}(2015){Ricker}, {Winn}, {Vanderspek}, {Latham},
  {Bakos}, {Bean}, {Berta-Thompson}, {Brown}, {Buchhave}, {Butler}, {Butler},
  {Chaplin}, {Charbonneau}, {Christensen-Dalsgaard}, {Clampin}, {Deming},
  {Doty}, {De Lee}, {Dressing}, {Dunham}, {Endl}, {Fressin}, {Ge}, {Henning},
  {Holman}, {Howard}, {Ida}, {Jenkins}, {Jernigan}, {Johnson}, {Kaltenegger},
  {Kawai}, {Kjeldsen}, {Laughlin}, {Levine}, {Lin}, {Lissauer}, {MacQueen},
  {Marcy}, {McCullough}, {Morton}, {Narita}, {Paegert}, {Palle}, {Pepe},
  {Pepper}, {Quirrenbach}, {Rinehart}, {Sasselov}, {Sato}, {Seager},
  {Sozzetti}, {Stassun}, {Sullivan}, {Szentgyorgyi}, {Torres}, {Udry}, \&
  {Villasenor}}]{tess15}
{Ricker}, G.~R., {Winn}, J.~N., {Vanderspek}, R., {et~al.} 2015, Journal of
  Astronomical Telescopes, Instruments, and Systems, 1, 014003,
  \dodoi{10.1117/1.JATIS.1.1.014003}

\bibitem[{{Roberts}(2013)}]{Roberts2013}
{Roberts}, M. S.~E. 2013, in IAU Symposium, Vol. 291, Neutron Stars and
  Pulsars: Challenges and Opportunities after 80 years, ed. J.~{van Leeuwen},
  127--132, \dodoi{10.1017/S174392131202337X}

\bibitem[{{Romani} \& {Sanchez}(2016)}]{rs16}
{Romani}, R.~W., \& {Sanchez}, N. 2016, \apj, 828, 7,
  \dodoi{10.3847/0004-637X/828/1/7}

\bibitem[{{Roy} {et~al.}(2015){Roy}, {Ray}, {Bhattacharyya}, {Stappers},
  {Chengalur}, {Deneva}, {Camilo}, {Johnson}, {Wolff}, {Hessels}, {Bassa},
  {Keane}, {Ferrara}, {Harding}, \& {Wood}}]{roy15}
{Roy}, J., {Ray}, P.~S., {Bhattacharyya}, B., {et~al.} 2015, \apjl, 800, L12,
  \dodoi{10.1088/2041-8205/800/1/L12}

\bibitem[{{Salvetti} {et~al.}(2017){Salvetti}, {Mignani}, {De Luca}, {Marelli},
  {Pallanca}, {Breeveld}, {H{\"u}semann}, {Belfiore}, {Becker}, \&
  {Greiner}}]{salvetti17}
{Salvetti}, D., {Mignani}, R.~P., {De Luca}, A., {et~al.} 2017, \mnras, 470,
  466, \dodoi{10.1093/mnras/stx1247}

\bibitem[{{Scargle} {et~al.}(2013){Scargle}, {Norris}, {Jackson}, \&
  {Chiang}}]{bayes}
{Scargle}, J.~D., {Norris}, J.~P., {Jackson}, B., \& {Chiang}, J. 2013, \apj,
  764, 167, \dodoi{10.1088/0004-637X/764/2/167}

\bibitem[{{Shahbaz} {et~al.}(2017){Shahbaz}, {Linares}, \&
  {Breton}}]{shahbaz17}
{Shahbaz}, T., {Linares}, M., \& {Breton}, R.~P. 2017, \mnras, 472, 4287,
  \dodoi{10.1093/mnras/stx2195}

\bibitem[{{Shahbaz} {et~al.}(2019){Shahbaz}, {Linares}, {Rodr{\'\i}guez-Gil},
  \& {Casares}}]{shahbaz19}
{Shahbaz}, T., {Linares}, M., {Rodr{\'\i}guez-Gil}, P., \& {Casares}, J. 2019,
  \mnras, 488, 198, \dodoi{10.1093/mnras/stz1652}

\bibitem[{{Stappers} {et~al.}(2014){Stappers}, {Archibald}, {Hessels}, {Bassa},
  {Bogdanov}, {Janssen}, {Kaspi}, {Lyne}, {Patruno}, {Tendulkar}, {Hill}, \&
  {Glanzman}}]{stappers14}
{Stappers}, B.~W., {Archibald}, A.~M., {Hessels}, J.~W.~T., {et~al.} 2014,
  \apj, 790, 39, \dodoi{10.1088/0004-637X/790/1/39}

\bibitem[{{Strader} {et~al.}(2014){Strader}, {Chomiuk}, {Sonbas}, {Sokolovsky},
  {Sand}, {Moskvitin}, \& {Cheung}}]{strader14}
{Strader}, J., {Chomiuk}, L., {Sonbas}, E., {et~al.} 2014, \apjl, 788, L27,
  \dodoi{10.1088/2041-8205/788/2/L27}

\bibitem[{{Strader} {et~al.}(2015){Strader}, {Chomiuk}, {Cheung}, {Sand },
  {Donato}, {Corbet}, {Koeppe}, {Edwards}, {Stevens}, {Petrov}, {Salinas},
  {Peacock}, {Finzell}, {Reichart}, \& {Haislip}}]{strader15}
{Strader}, J., {Chomiuk}, L., {Cheung}, C.~C., {et~al.} 2015, \apjl, 804, L12,
  \dodoi{10.1088/2041-8205/804/1/L12}

\bibitem[{{Swihart} {et~al.}(2019){Swihart}, {Strader}, {Chomiuk}, \&
  {Shishkovsky}}]{swihart19}
{Swihart}, S.~J., {Strader}, J., {Chomiuk}, L., \& {Shishkovsky}, L. 2019,
  \apj, 876, 8, \dodoi{10.3847/1538-4357/ab125e}

\bibitem[{{Swihart} {et~al.}(2018){Swihart}, {Strader}, {Shishkovsky},
  {Chomiuk}, {Bahramian}, {Heinke}, {Miller-Jones}, {Edwards}, \&
  {Cheung}}]{swihart18}
{Swihart}, S.~J., {Strader}, J., {Shishkovsky}, L., {et~al.} 2018, \apj, 866,
  83, \dodoi{10.3847/1538-4357/aadcab}

\bibitem[{{Szkody} {et~al.}(2003){Szkody}, {Fraser}, {Silvestri}, {Henden},
  {Anderson}, {Frith}, {Lawton}, {Owens}, {Raymond}, {Schmidt}, {Wolfe},
  {Bochanski}, {Covey}, {Harris}, {Hawley}, {Knapp}, {Margon}, {Voges},
  {Walkowicz}, {Brinkmann}, \& {Lamb}}]{szkody03}
{Szkody}, P., {Fraser}, O., {Silvestri}, N., {et~al.} 2003, \aj, 126, 1499,
  \dodoi{10.1086/377346}

\bibitem[{{Takata} {et~al.}(2014){Takata}, {Li}, {Leung}, {Kong}, {Tam}, {Hui},
  {Wu}, {Xing}, {Cao}, {Tang}, {Wang}, \& {Cheng}}]{takata14}
{Takata}, J., {Li}, K.~L., {Leung}, G. C.~K., {et~al.} 2014, \apj, 785, 131,
  \dodoi{10.1088/0004-637X/785/2/131}

\bibitem[{{Tam} {et~al.}(2010){Tam}, {Hui}, {Huang}, {Kong}, {Takata}, {Lin},
  {Yang}, {Cheng}, \& {Taam}}]{tam10}
{Tam}, P.~H.~T., {Hui}, C.~Y., {Huang}, R.~H.~H., {et~al.} 2010, \apjl, 724,
  L207, \dodoi{10.1088/2041-8205/724/2/L207}

\bibitem[{{Thorstensen} \& {Armstrong}(2005)}]{ta05}
{Thorstensen}, J.~R., \& {Armstrong}, E. 2005, \aj, 130, 759,
  \dodoi{10.1086/431326}

\bibitem[{{van der Walt} {et~al.}(2011){van der Walt}, {Colbert}, \&
  {Varoquaux}}]{numpy11}
{van der Walt}, S., {Colbert}, S.~C., \& {Varoquaux}, G. 2011, Computing in
  Science Engineering, 13, 22, \dodoi{10.1109/MCSE.2011.37}

\bibitem[{{van Staden} \& {Antoniadis}(2016)}]{va16}
{van Staden}, A.~D., \& {Antoniadis}, J. 2016, \apjl, 833, L12,
  \dodoi{10.3847/2041-8213/833/1/L12}

\bibitem[{{Virtanen} {et~al.}(2019){Virtanen}, {Gommers}, {Burovski},
  {Oliphant}, {Cournapeau}, {Weckesser}, {alexbrc}, {Peterson}, {endolith},
  {Mayorov}, {van der Walt}, {Wilson}, {Laxalde}, {Brett}, {Millman}, {Lars},
  {Nelson}, {Haberland}, {eric-jones}, {Polat}, {Larson}, {Kern}, {Moore},
  {Carey}, {Leslie}, {Perktold}, {Reddy}, {Bharti}, {Feng}, \&
  {Vanderplas}}]{scipy19}
{Virtanen}, P., {Gommers}, R., {Burovski}, E., {et~al.} 2019, {scipy/scipy:
  SciPy 1.2.1}, v1.2.1,  Zenodo, \dodoi{10.5281/zenodo.2560881}

\bibitem[{{Woudt} {et~al.}(2004){Woudt}, {Warner}, \& {Pretorius}}]{woudt04}
{Woudt}, P.~A., {Warner}, B., \& {Pretorius}, M.~L. 2004, \mnras, 351, 1015,
  \dodoi{10.1111/j.1365-2966.2004.07843.x}

\end{thebibliography}
\bibliographystyle{aasjournal}

\end{document}